# Causality and time-space evolution of physical systems.

Enrique Ordaz Romay [1]

*Facultad de Ciencias Físicas, Universidad Complutense de Madrid*

## Abstract

All physical process are subject to some laws which determine with math accurately its time-space evolution. These laws are described, in the last analysis for the principle of causality. The physical space can be homogeneous or heterogeneous in function of the distribution of the physical magnitudes. Applying the principle of causality to the homogeneous space we observe that its more important qualities are: their metric does not depend of the 4coordeninate where it is measured and the light speed constancy [1]. However, to apply the principle of causality in the heterogeneous space is very much fruitful. Thanks to the presence of physical magnitudes that vary in the 4space we can find the equations that this magnitudes should verify so that it takes place their space and time evolution.

Working on the principle of causality like axiom for describe our physical systems, we will obtain the most basic expression of evolution of the magnitudes of the physical system. This expression coincides with the principle of action expressed for a tensorial action [2]. The scalar form of this principle coincides, in its turn, with the traditional relativistic and quantum expression, of the principle of action or principle of least action according to other authors.

---

[1] enorgazro@cofis.es

# Introduction.

The whole science is based on an unique principle well-known as the principle of causality [3]. This principle can enunciate as: "all event comes determined by the history of previous events. To these previous events we called causes and to the event that determine we called effect." This principle is so essential because "it is verify independently of the reference frame that we use." To apply this principle to the homogeneous systems forces us, for not breaking the homogeneity of the system, to consider the causes and the effects like external events to our system.

For this reason, the unique consequence that we can deduce from the principle of causality applied to the homogeneous spaces is that one deduces from the propagation of the sign (luminous).

However, in the heterogeneous space we can apply all the power of the principle of causality and in this form to obtain the equations of space-time evolution of these systems.

# The heterogeneous system.

It is defined the heterogeneous system as that physical system which has characteristic differentiating from some regions to others. Such characteristics are due to the existence of physical magnitudes whose values varies in function of the 4-coordinateses in which are measured.

Since, when a physical system is heterogeneous it is not homogeneous, then its metric, cannot be constant in all the points of the system, that is to say, the metric of a heterogeneous system is variable with the coordinateses of the 4point in which we measure it: $G = G(X)$.

Although the metric is not constant, for the properties of the distance, it continues being symmetrical. By virtue of this symmetry we always will be able to find for each point of the system a reference frame, together with a units system in the one which, the metric is similar to $G_0$. This principle is known like principle of strong equivalence.

## The concept of physical magnitude.

A physical magnitude is a quality of (or in) the 4space that it is characterized by to be comparable and scalable with another quality of the same type to which we call pattern.

The fact of observe a physical magnitude in our system assures us that the system is not homogeneous. Any hypothesis of homogeneity on our system would be annulled as soon as we could detect the presence of a physical magnitude, since this, would characterize at the 4region of the space where it showed, in front of the rest of the regions of the system.

The concept of physical magnitude is an abstraction that traditionally has united to the concept of matter and not to the space. This way, the mass or the electrical charge are considered physical magnitudes, while the electric, magnetic or gravitational fields are considered, rather, qualities of the space. However, these fields have units systems, and places of the space in which are measured their values. Therefore, we will also include them in the concept of physical magnitude.

In this form, the physical magnitude will adopt the most general form of its definition: "quality of (or in) the 4space that it is characterized by to be comparable and scalable with another quality of the same type to which we call pattern."

The physical magnitudes have, by virtue of their definition, a mathematical representation. This mathematical representation can make in many forms and in fact it has gone changing along the history according to the necessities of the mathematical apparatus that it has used in each time. Today in day the mass or the electrical charge are considered like a to scalar (tensor of order 0), the electrical field strength is a vector (tensor of 1º order), the electromagnetic field is a matrix (tensor of 2º order) or the gravitational field like a tensor of 4º order [4].

In a general form we can say that, using the differential geometry, the present physical magnitude in a point of the system can be represented by a mathematical object of range *n* that is applied in a system $\Omega$ [5]. That is to say: *P* is a physical magnitude if

$$P : \Omega^* \times \Omega^* \times ... \times \Omega^* \to \Re \; ; \; P = p_{ijk...} dx^i \times dx^j \times dx^k \times ...$$

The particular realization that is made of the physical magnitudes will depend in each case of the purpose. In a purely mechanical analysis of material points the magnitudes that will interest us will be the masses and speeds of each one of the particles of the system. These magnitudes are carried out in the concept of lineal moment, being $m_i$ the mass and $v_i$ the speed of the *i* particle:

$$m_i \cdot \vec{v}_i$$

Equally, for an electrodynamic system, the magnitude would be carried out in the next form:

$$m(X) \cdot \vec{v}(X) - e(X)\vec{A}(X)$$

Being *m* the mass distribution, *v* the speed, *e* electric distribution of charge and *A* the gauge potential of the electromagnetic field.

In the analysis that we will continue hereinafter, we will do without the mathematical representation of the physical magnitudes and we will centre in looking for that mathematical expression describes to our physical system.

## Principle of causality and physical magnitudes.

The principle of causality, just as it is enunciated in any dictionary, says "all event comes determined by the history of previous events. To these previous events we called causes and to the event that they determine we called effect."

This principle can to be applicable to the physics with some considerations:

An event is a circumstance that characterizes a point or region of the 4space, differing of its neighbourhood.

Considering the definitions of heterogeneous space and physical magnitude, it is clear that an event, what manifests is a physical magnitude that differentiates a region of the 4space.

In these circumstances, the principle of causality is translated by: "all physical magnitude comes determined by the history of previous physical magnitudes."

Since the physical magnitudes are represented by mathematical objects, this determinism is translated in mathematical expressions that connect the physical magnitudes "cause" with the physical magnitudes "effects".

## Evolution of a physical system.

We begin, then, from a physical system in which there are some physical magnitudes represented by a mathematical object of degree *n* to which we represent by *P(X)*.

We will call evolution of our physical system to the mathematical expression that shows us how vary the physical magnitudes in function of the 4coordinates, inside the system.

The variation of the magnitudes, in function of the dimensions, when these they are discreet, is totalled according to the expression: $\sum_{\forall X \in \Omega} P(X)$. Similarly for continuous dimensions we total according to $\lim_{\Delta X \to 0} \sum_{\forall X \in \Omega} P(X) \Delta X$ that coincides with the definition of the integral function. In this case *P* represent a distribution density of the magnitudes in function of the 4coordinates of the system.

According to this, the evolution of a physical system $\Omega$ comes determined by a $n-1$ differential form [5] that is defined as:

$$S = \int_\Omega P \tag{1}$$

that is to say, the magnitudes represented by $P$ verify:

$$P_{\mu ij\ldots} = \frac{1}{(n-1)!} \partial_{[\mu} S_{ij\ldots]} \tag{2}$$

A more compact notation of this expression would be, using the differential geometry:

$$\frac{1}{(n-1)!} \partial_{[\mu} S_{ij\ldots]} = D_\mu S_{ij\ldots}$$

The expressions (1) and (2), if we work in the classic space, in Cartesian coordinates, we can reduce them to a more traditional form:

$$S = \int_\Omega P = \int_{X_1}^{X_2} P(X)dX = \int_1^2 \left(\iiint_{2\to 1} P\,dxdydz\right) cdt$$

When $P$ is a vector, $S$ will be a scalar. Then, the expression (2) will take the form:

$$P^i(X) = \frac{\partial S}{\partial x^i}$$

These last two expressions indicate us that, in the classic space, in Cartesian coordinates and being $P$ a vector and therefore $S$ a scalar, $S$ coincides with the expression of the action function when this is function of the coordinateses and $P$ with the moment of the system.

This fact is not fortuitous. We have began from that *P* is the representation of the magnitudes of the system and *S* is the evolution of the system. On the other hand, the action, just as it describes by the analytic physics, is the mathematical expression that describes to a physical system and the moment of the system is which includes the physical magnitudes and as their interactions.

However, in our analysis, we have not supposed the existence of a concrete realization for the mathematical representation of the magnitudes, but rather we have began from the supposition of this representation corresponds with a mathematical *n*-object and this supposes an essential difference. The action and the moment that we present do not have necessarily to be a scalar and a vector, but rather, in general they will be a (*n*-1)-cotensor and a mathematical *n*-object respectively.

Therefore, we will call to *S* tensorial action and *P* tensorial moment. To the scalar and vectorial magnitudes that traditionally represent to the action and the moment, we will represent them with a line underlined to differentiate them.

## Lorentz scalars and action principle.

We are interested in knowing that mathematical expression should verify the tensorial action for check the principle of causality.

As we have already seen, the direct application of the principle of causality is translated in the expression:

$$\boxed{S = \int_{\tilde{\upsilon}} P}$$

This form of express the tensorial action assures us that *S* verifies the properties of the tensors [5], or in the scalar case, the properties of a Lorentz scalar. As we know, it is said that a scalar is a scalar of Lorentz when it doesn't depend of the coordinated frame in which is expressed.

The scalar product of the tensor action for itself is a scalar of Lorentz too, because in a change of coordinated from $X$ to $X'$:

$$S^2 = S^{ijk...} S_{ijk...} = S'^{ijk...} \frac{\partial X'}{\partial X} \frac{\partial X}{\partial X'} S'_{ijk...} = S'^{ijk...} S'_{ijk...} = S'^2$$

On the other hand, if $\underline{f}(X)$ is a scalar function of Lorentz, then any transformation in the coordinateses maintains unaffected the expression. That is to say $\underline{f}(X) = \underline{f}(X + \delta X)$. Therefore, if $\underline{f}(X)$ is a scalar function of Lorentz then, a variation in this function is equal to zero. Mathematically:

$$\delta \underline{f}(X) = \underline{f}(X) - \underline{f}(X + \delta X) = 0$$

Applying this property to $S^2$, we obtain the expression:

$$\boxed{\delta(S^2) = 0} \qquad (3)$$

This expression is the relationship that should complete the tensorial action to be the mathematical representation of the evolution of a physical system. For similarity with the case of the scalar action, to the equation (3) we will call it principle of tensorial action, principle of extreme tensorial action or principle of least tensorial action.

It is easy to find the scalar equivalent expression equation (3). Since it is a to $S^2$, we can define the scalar $\underline{S} = +\sqrt{S^2}$. According to this, the property $\delta(S^2) = 0$ is equivalent to $\delta(\underline{S}^2) = 0$ that you can develop from a scalar form until the expression: $\delta(\underline{S}^2) = 0 \Rightarrow \underline{S} \cdot \delta \underline{S} = 0 \Rightarrow \underline{S} = 0$ ó $\delta \underline{S} = 0$.

Since both $\underline{S}$ and $S$ are defined except in a constant, the condition $\underline{S} = 0$ is arbitrary. That is to say: In all physical systems should be verify the condition $\delta \underline{S} = 0$. As this is the condition of the action principle of Hamilton [6], this justifies us that we call to $\underline{S}$ scalar function action and to $S$ tensorial function action or simply action.

It is trivial to observe that, when the mathematical realization of *P* is vectorial, then the traditional action coincides with <u>S</u>. For this reason, the principle of Hamilton is a particularization of the equation (3).

In consequence: all physical systems comes determined by a tensorial function *S*, to which we will call action, that verify the principle of action expressed by: $\delta(S^2) = 0$. As particularization we can say that all physical systems has defined a scalar function <u>S</u> that it verify the condition of minimal action: $\delta(\underline{S}) = 0$.

# References

[1] E. Ordaz, *Causality, light speed constancy and local action principle*. (2002)

[2] E. Ordaz, *Tensorial origin of the action function, nexus between Quantum physics and General Relativity*. (2002)

[3] B. Russell, *An outline of philosophy*, (1936).

[4] ] L. D. Landau y E. M. Lifshitz, *The Classical Theory of fields*, 2ª ed., (1967)

[5] B. Felsager, Geometry, particles and fields. (1981)

[6] L. D. Landau and E. M. Lifshitz, Mechanics, (1965)